\documentclass[journal=jacsat,manuscript=article]{achemso}
\usepackage[version=3]{mhchem}

\usepackage{color}
\usepackage{amssymb}

\DeclareUnicodeCharacter{2009}{\,}
\DeclareUnicodeCharacter{2212}{-}
\author{Tanima Kundu}
\affiliation{School of Physical Sciences, Indian Association for the Cultivation of Science, 2A $\&$ B
Raja S. C. Mullick Road, Jadavpur, Kolkata - 700032, India
}

\author{Barnik Pal}
\affiliation{School of Physical Sciences, Indian Association for the Cultivation of Science, 2A $\&$ B
Raja S. C. Mullick Road, Jadavpur, Kolkata - 700032, India
}

\author{Bikash Das}
\affiliation{School of Physical Sciences, Indian Association for the Cultivation of Science, 2A $\&$ B
Raja S. C. Mullick Road, Jadavpur, Kolkata - 700032, India
}

\author{Rahul Paramanik}
\affiliation{School of Physical Sciences, Indian Association for the Cultivation of Science, 2A $\&$ B
Raja S. C. Mullick Road, Jadavpur, Kolkata - 700032, India
}

\author{Sujan Maity}
\affiliation{School of Physical Sciences, Indian Association for the Cultivation of Science, 2A $\&$ B
Raja S. C. Mullick Road, Jadavpur, Kolkata - 700032, India
}

\author{Anudeepa Ghosh}
\affiliation{School of Physical Sciences, Indian Association for the Cultivation of Science, 2A $\&$ B
Raja S. C. Mullick Road, Jadavpur, Kolkata - 700032, India
}

\author{Mainak Palit}
\affiliation{School of Physical Sciences, Indian Association for the Cultivation of Science, 2A $\&$ B
Raja S. C. Mullick Road, Jadavpur, Kolkata - 700032, India
}

\author{Marek Kopciuszynski}
\affiliation{Sincrotrone Trieste s.c.p.a., 34149 Basovizza, Trieste, Italy}

\author{Alexei Barinov}
\affiliation{Sincrotrone Trieste s.c.p.a., 34149 Basovizza, Trieste, Italy}

\author{Sanjoy Kr Mahatha} 
\affiliation{UGC-DAE Consortium for Scientific Research, Khandwa Road, Indore 452001, Madhya Pradesh, India}

\author{Subhadeep Datta}
\email{sspsdd@iacs.res.in}
\affiliation{School of Physical Sciences, Indian Association for the Cultivation of Science, 2A $\&$ B
Raja S. C. Mullick Road, Jadavpur, Kolkata - 700032, India
}

\title[An \textsf{achemso} demo]{Tunable Electron Transport in Defect-Engineered PdSe$_2$}

\makeatother
\begin{document}

\begin{abstract}

Tuning the ambipolar behavior in charge carrier transport \textit{via} defect-engineering is crucial for achieving high mobility transistors for nonlinear logic circuits. Here, we present the electric-field tunable electron and hole transport in a microchannel device consisting of highly air-stable van der Waals (vdW) noble metal dichalcogenide (NMDC), PdSe$_2$, as an active layer. Pristine bulk PdSe$_2$ constitutes Se surface vacancy defects created during the growth or exfoliation process and offers an ambipolar transfer characteristics with a slight electron dominance recorded in field-effect transistor (FET) characteristics showing an ON/OFF ratio $<$ 10 and electron mobility $\sim$ 21 cm$^2$/V.s. However, transfer characteristics of PdSe$_2$ can be tuned to a hole-dominated transport while using hydrochloric acid (HCl) as a \emph{p}-type dopant. On the other hand, the chelating agent EDTA, being a strong electron donor, enhances the electron-dominance in PdSe$_2$ channel. In addition, \emph{p}-type behavior with a 100 times higher ON/OFF ratio is obtained while cooling the sample down to 10 K. Low-temperature angle-resolved photoemission spectroscopy resembles the \emph{p}-type band structure of PdSe$_2$ single crystal. Also, first principle density functional theory calculations justify the tunability observed in PdSe$_2$ as a result of defect-engineering. Such a defect-sensitive ambipolar vdW architecture may open up new possibilities towards future CMOS (Complementary Metal-Oxide-Semiconductor) device fabrications and high performance integrated circuits.  

\end{abstract}

\maketitle

\section{Introduction}
 
Performance of a FET based on downscaled two-dimensional (2D) vdW semiconductors like transition metal dichalcogenides (TMDCs) (\textit{e.g.} MoS$_2$, WS$_2$, MoSe$_2$ etc.) as channel depends on the layer's chemical and environmental stability \cite{WS2,MoS2,MoSe2}. In TMDCs, aging effects like gradual oxidation at the TM edges of a micromechanically exfoliated flake transferred from the bulk crystal or at the grain boundaries of a large area sheet hinders its enduring applications in nanoelectronics and optoelectronics. As it happens, easy dissociation of molecular oxygen at the TM edge (\textit{e.g.} Mo edge in MoS$_2$) with low barrier energy (close to 0.3 eV) compared to its protection at the surface (1.5 eV on the MoS$_2$ surface) makes the layer more susceptible to wrinkle formation, morphological variation with time. Nonetheless, for the fundamental research concerning electronic/optoelectronic devices, utilizing the standard crystal-growth techniques like chemical vapor deposition (CVD) or chemical vapor transport (CVT) \cite{CVT,MoS2_CVD}, TMDC flakes/nanostructures can be grown on the conventional SiO$_2$/Si substrate and contacted with metal electrodes using lithography, followed by electron/hole transport measurements in the FET geometry depending on the band structure and doping profile. Also, highly disparate atomic layers can be stacked together, \textit{e.g.} graphene and MoS$_2$, to form vdW heterostructures without any constraint of lattice matching \cite{gr-hBN,gr-MoS2}. Surface charge transfer using donor/acceptor molecule for suitable doping in few-layer TMDC, like WSe$_2$, may result in improved ON/OFF ratio and higher mobility than the pristine samples \cite{WSe2-HCl}. Besides, hydracid treatment in monolayer MoSe$_2$ reduces the Se-vacancies by passivation, and thereby, enhances the photoluminescence \cite{MoSe2-HBr}. Similarly, using ethylenediaminetetra acetic acid (EDTA) as electron donor, repairing of single/double Se vacancies can be achieved in single layer of MoSe$_2$, and as a result of which several times of increment in the carrier density can be observed \cite{MoSe2-EDTA}.               

Although notable advancement has been perceived in the device electronics with defect-engineered  surfaces of TMDC, researchers got involved in exploring a new family of 2D vdW semiconducting materials, noble metal dichalcogenides (NMDCs) with the stoichiometry $MX_2$ (metal M = Pt, Pd, and chalcogen X = S, Se, Te) which may provide superior air-stability with similar or enhanced FET performance \cite{review_1}. In contrast to the conventional Group 6 TMDCs, NMDCs with group 10 elements possess largely occupied \emph{d}-orbitals (more than half-filled) of the noble metals and highly hybridized \emph{p$_z$} orbitals of the chalcogen atoms which may lead to strong interlayer coupling as well as layer dependent  band-gap, as observed in one of the compound from NMDC family, PtS$_2$ ($\sim$0.2 eV in bulk and $\sim$1.6 eV in the monolayer) \cite{PtSe2,PdSe2,PtS2}. Additionally, it has been predicted that coordinatively unsaturated noble metal, like Pd, will have much higher O$_2$ adsorption energy barrier ($\sim$ 1.2 eV) than Group 6 elements which might be the reason for high air-stability at the edges \cite{PdSe2,PtSe2_2}. Recent experimental study has confirmed that the device performance of PtSe$_2$ FET remains nearly unchanged after 5 months of air-exposure \cite{PtSe2_2}. Apart from the air-stability and layer dependence, the electronic properties of these materials vary extensively due to surface vacancy defects introduced during the growth and exfoliation process or can be further tuned by doping \cite{prb-defect,npj-defect,advm-defect,PtSe2-defect}. Altogether, air-stable NMDCs in the reduced dimension may exhibit better device performance and tunability \textit{via} defect engineering  for future applications in logic circuits \cite{CMOS}. 

Recently, palladium diselenide (PdSe$_2$), a Group 10 NMDC has been explored and recognized as a highly air-stable layered material with a unique puckered pentagonal morphology \cite{PdSe2}. Usually, puckered configuration exhibits novel anisotropic properties which is rarely found in the family of 2D materials. Unlike the conventional TMDCs and other NMDCs, in the unit cell of PdSe$_2$, each Pd atom is surrounded by four Se-Se intralayer dimers and thereby forms a PdSe$_4$ unit. This type of structure makes the crystal more defect-sensitive due to the kinetic barrier of the Se-vacancy diffusion which is much lower than any other TMDC resulting in a structural distortion by the symmetry breaking of Se-Se intralayer bonds \cite{diffusion-1,diffusion-2}. This defect-sensitivity potentially enriches electronic transport as well as phase change engineering. Previously, first-principle calculations predicted the tunability of layered PdSe$_2$ due to surface vacancy defects \cite{prb-defect,sci-rep-defect}. But the experimental regime of defect engineering in PdSe$_2$ is still unexplored. 
\newline
Here, we report tunable electronic transport properties of defect-engineered bulk layer of vdW material PdSe$_2$. Being a low band gap semiconductor in pristine bulk (E$_g$ $\sim$ 0.3 - 0.4 eV), PdSe$_2$ shows consistent ambipolar transfer characteristics with ON/OFF ratio $<$ 10 and electron mobility $\sim$ 21 cm$^2$/V.s in the ambient condition while using as channel in the FET. The mobility value is quite large in comparison to TMDCs like MoS$_2$ \cite{MoS2-mobility}. By lowering the temperature the ON/OFF ratio gets improved more than 100 times as the contribution of defect states is reduced and a \emph{p}-type semiconducting transport has been observed, evidenced by low temperature angle-resolved photoemission spectroscopy (ARPES) measurement, where the valence band maxima touches the Fermi surface at the $\Gamma$ point.
Further, we have controlled the Se surface defects \textit{via} processing the pristine devices with \emph{p}-type dopant as hydrochloric acid and \emph{n}-type dopant as EDTA disodium salt (EDTA/Na$_2$) solution. Transfer characteristics of the solution-processed modified flakes backed up by the first-principle density functional theory (DFT) calculations confirms the effect of doping which allows to change the carrier concentration based on the charge transfer between the dopants (hole from Cl and electron from EDTA) and Pd-ions around the Se-vacancies and thereby, modulates the effective barrier height at the metal-semiconductor junction. Tuning ambipolar behavior with doping may offer potential devices for field-programmable \emph{p-n} junctions in future microelectronics beyond silicon. 

\section{Experimental details}
\label{subsec:exp}
{\bfseries Synthesis and Characterization:} Standard self-flux method was adopted to grow bulk PdSe$_2$ single crystals as referenced by Akinola $et\ al$ \cite{PdSe2}. Pd (99.95\% , Alfa Aesar) and Se (99.99\% , Alfa Aesar) powders were mixed thoroughly in an atomic ratio 1:6. The mixture was then loaded into a quartz tube followed by evacuating the tube under a vacuum condition 10$^{-5}$ Torr, sealed and then placed in a single-zone furnace. The furnace temperature was raised to 850$^{\circ}$C at a rate of 1$^{\circ}$C/min. After holding 50 hours at this temperature, the furnace was allowed to cool slowly to 450$^{\circ}$C at a rate of 3$^{\circ}$C/h and then to room temperature. PdSe$_2$ single crystals (Figure S1a) were extracted by cleaving the chunk obtained inside the quartz tube. The bulk flakes were transferred on a clean SiO$_2$/Si substrate $via$ scotch-tape or PDMS by mechanical exfoliation technique. Also liquid phase exfoliation was performed in isopropyl alcohol by sonicating the solution using a bath sonicator.

Next, single crystal X-ray diffraction (SXRD) was performed using Bruker APEX II (CCD area detector, Mo K$\alpha$, $\lambda = 0.7107$ {\AA}) at room temperature followed by the structure determination and refinement using the corresponding software packages of the diffractometer. Transmission electron microscope (TEM) characterization was carried out in a JEOL-JEM-F200 electron microscope by giving an accelerating voltage 200 kV. Scanning electron microscope (SEM) and energy-dispersive spectroscopic (EDS) study were also performed in a JEOL JSM-6010LA. Atomic force microscope (AFM) (Asylum Research MFP-3D) was used to determine the thickness of the PdSe$_2$ flakes. Raman spectroscopy was performed for optical characterization using a Horiba T64000 Raman spectrometer. X-ray photoelectron spectroscopy (XPS) was carried out using an OMICRON electron
spectrometer (model no. XM 1000).

Subsequently, angle-resolved photoemission spectroscopy (ARPES) measurements were performed on a freshly cleaved PdSe$_2$ crystal in ultrahigh vacuum at the Spectromicroscopy beamline of the Elettra Sincrotrone in Trieste, Italy. The spectra were acquired at 95 K with a photon energy of 74 eV. The linearly polarized incident radiation was focused to a spot size of approximately 1 $\mu$m by a Schwarzschild objective \cite{arpes1}. The scanning stage of the sample was used for positioning and raster imaging the sample with respect to the fixed photon beam. Photoemission intensity maps were taken by rotating the hemispherical analyzer mounted on a two-axis goniometer. The total energy and angular resolutions were better than 50 meV and 0.35$^{\circ}$, respectively.  

{\bfseries Device Fabrication and electrical characterization:} After mechanical exfoliation on a clean SiO$_2$ (285 nm)/Si substrate, suitable flakes were located using optical microscope. Then optical lithography was used for device fabrication. Metal contacts with 10 nm Cr followed by 80 nm Au were made using a thermal evaporator. For the four-probe measurement the device was protected with a few-layer hexagonal boron nitride (hBN) flake by micro-manipulation technique. The FET measurements were carried out using Keithley 2450 and 2601B source measure units (SMU) and temperature dependent transport studies were performed by a closed cycle cryostat (ARS-4HW). For the \emph{p}-type doping, PdSe$_2$ devices were dipped into 11M HCl solution for 10 mins and then rinsed with deionized water. On the other hand, for \emph{n}-type doping, PdSe$_2$ devices were processed by a drop of EDTA/Na$_2$ solution which was prepared by mixing 0.2M EDTA disodium salt in deionized water and its pH scale was adjusted to 8.0 by adding a few NaOH pellets in the solution. The solution-processed devices were kept in a vacuum desiccator overnight to dry off. Three-terminal carrier transport measurements of the doped devices were carried out following the above mentioned set-ups.   

  


\section{Computational details}
\label{subsec:theory}

Density functional theory (DFT) calculations were carried out using the Vienna Ab initio Simulation Package (VASP) \cite{VASP1,VASP2}. Projector-augmented-wave (PAW) pseudopotential was used to incorporate electron-ion interaction. Generalized gradient approximation (GGA) in the form of Perdew-Burke-Ernzerhof (PBE) \cite{PBE} was considered to describe the exchange-correlation effects. The plane wave cut-off energy was set to 380 eV for pristine, vacancy defect induced and Cl-doped PdSe$_2$ and 520 eV for EDTA-doped PdSe$_2$. The convergence criteria for interatomic forces and electronic minimizations were set to 0.001 eV/{\AA} and $10^{-6}$ eV respectively. To simulate the defects, all the calculations were performed for a $2\times 2\times 2$ supercell with $4\times 4\times 2$ and $8\times 8\times 6$ k-point sampling generated according to the Monkhorst-Pack scheme \cite{MP-scheme} for structure relaxations and density of states (DOS) calculations, respectively. The phonon density of states and band structure calculations were carried out using the PHONOPY \cite{PHONOPY} software. Raman spectrum was simulated using Quantum Espresso DFT package \cite{QE1,QE2}. To visualize the crystal structures VESTA software \cite{VESTA} was used.

\section{Results and discussion}
\label{sec:results_discussions}

\begin{figure}[ht]
\centerline{\includegraphics[scale=0.8, clip]{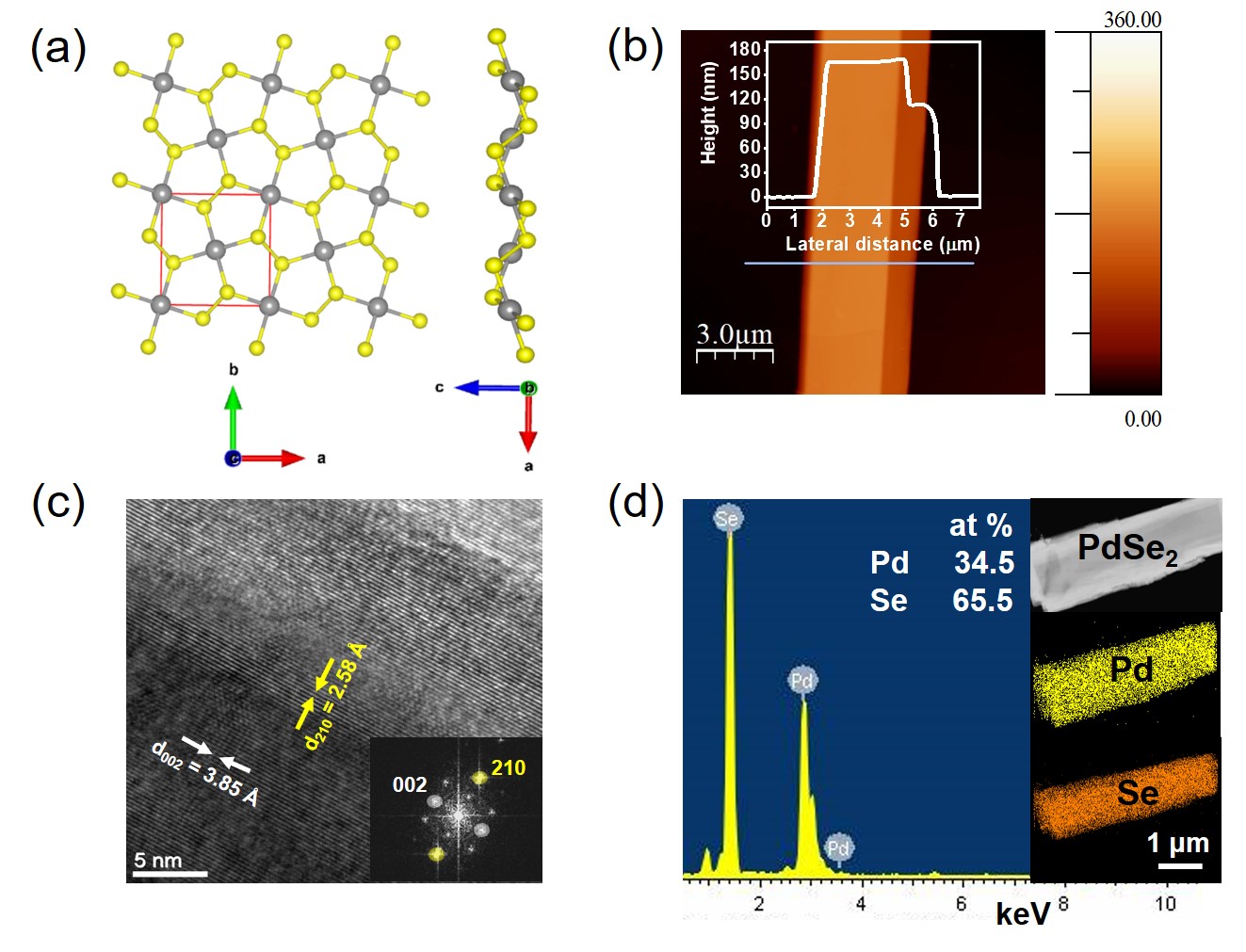}}
\caption{Palladium diselenide (PdSe$_2$): (a) Top and side view of the orthorhombic crystal structure of PdSe$_2$ in a puckered pentagonal morphology. The unit cell is represented by the red box. (b) AFM image with the corresponding height profile of a bulk PdSe$_2$ flake exfoliated on a SiO$_2$/Si substrate. (c) TEM characterization showing (002) and (210) lattice planes with the corresponding interplanar spacing 3.85 \AA $ $ and 2.58 \AA, respectively. Inset shows the associated SAED pattern with the different bright spots representing different lattice planes. (d) Elemental mapping of an exfoliated bulk PdSe$_2$ flake with 34.5\% of Pd and 65.5\% of Se. 
\label{char}}
\end{figure}

Using the SXRD refinement, the structure of the as-grown crystals was calculated to be orthorhombic $Pbca$ (as shown in Figure \ref{char}a) with the lattice parameters $a=5.77$ {\AA}, $b=5.91$ {\AA}, $c=7.73$ {\AA} and $\alpha=\beta=\gamma=90^{\circ}$ (point group symmetry $D_2h$). Layers of PdSe$_2$ are stacked together with an interlayer vdW force along the $c$-axis. The unit cell consists of four Pd atoms and eight Se atoms in a unique puckered pentagonal morphology. Each Pd atom is tetra-coordinated with Se atoms, due to the typical $d^8$ configuration of Pd$^{2+}$ ions which usually adopts square-planar coordination. 

\begin{figure}[ht]
\centerline{\includegraphics[scale=0.85, clip]{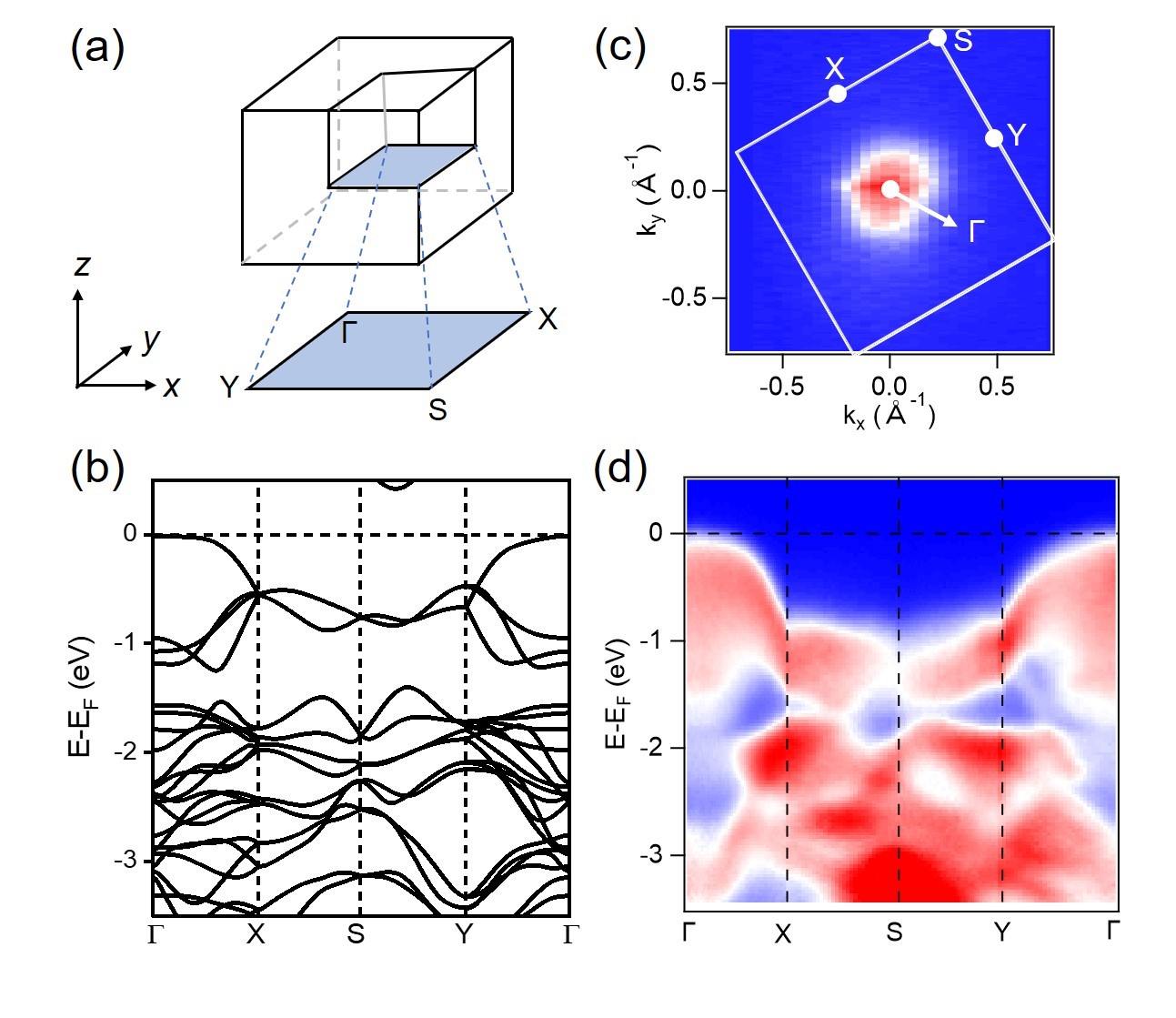}}
\caption{(a) 3D and a quadrant of projected 2D surface Brillouin zone with high symmetry points. (b) DFT-calculated band structure of bulk pristine PdSe$_2$ along the high symmetry directions. ARPES spectra of \textit{in-situ} cleaved PdSe$_2$ with its (c) Fermi surface and (d) band dispersions along the high symmetry directions.
\label{arpes}}
\end{figure}

Mechanical exfoliation was performed using a scotch-tape or PDMS to transfer PdSe$_2$ flakes on Si substrate coated with an SiO$_2$ layer of width 285 nm, whereas liquid phase exfoliation was carried out in isopropyl alcohol. The height profile shown in Figure \ref{char}b corresponding to the atomic force microscope (AFM) image confirmed the bulk nature of the chosen flakes (thickness $\sim$ 150 nm). FESEM image of a PdSe$_2$ single crystal is shown in Figure S1b. EDS study confirmed the desired stoichiometry of the as-grown crystals (33.24\% of Pd and 66.76\% of Se). Core level X-ray photoelectron spectra for Pd $3d$ and Se $3d$ orbitals are shown in Figure S1c. Pd $3d_{5/2}$ and $3d_{3/2}$ peaks are located at 336.1 eV and 341.3 eV, respectively and Se $3d_{5/2}$ and $3d_{3/2}$ peaks are located at 54.1 eV and 54.8 eV, respectively which are consistent with the reported binding energies of PdSe$_2$ crystals \cite{adfm-nishiyama}. The lattice planes of PdSe$_2$ crystal were characterized by Transmission Electron Microscope (TEM) as shown in Figure \ref{char}c. The d-spacings were calculated to be 3.85 \AA $ $ and 2.58 \AA $ $ for planes (002) and (210), respectively which are consistent with that obtained from the crystallographic information file (CIF) extracted from SXRD refinement. Inset of Figure \ref{char}c displays the selected area electron diffraction (SAED) pattern where the first order bright spots denote different Bragg planes. The elemental mapping of the exfoliated flake is depicted in Figure \ref{char}d which indicates that a slight Se vacancy created at the time of exfoliation (34.5 \% of Pd and 65.5 \% of Se). Raman spectroscopy for pristine as well as doped samples was carried out for optical characterization which exhibits six vibrational modes (3A$_g$ and 3B$_{1g}$) including a mixed one (see supplementary for details). 
\newline
\begin{center}
\bfseries{A. Angle-resolved photoemission spectroscopy}\\
\end{center}
To investigate the electronic structure of pristine bulk PdSe$_2$ crystal and confirm the electronic homogeneity, angle-resolved photoemission spectroscopy (ARPES) experiment  were performed with a $\mu$-focused photon beam. Figure \ref{arpes}b shows the simulated band structure along the high symmetry directions as illustrated in the Brillouin zone of bulk PdSe$_2$ in Figure \ref{arpes}a. As can be seen, the valence band maximum (VBM) is located at the $\Gamma$ point, whereas, the conduction band minimum (CBM) appears along the SY symmetry direction. The experimental band structure of this material along the high symmetry directions shown in Figure \ref{arpes}d is further corroborated using DFT calculations (see Figure \ref{arpes}b) and previous ARPES reports \cite{arpes2,arpes3,arpes4}. This also substantiates the single-crystallinity of the grown crystals. Figure \ref{arpes}c shows the Fermi surface map where a weak spectral weight originating from the VBM at $\Gamma$ is visible without any additional feature originating from the CBM along the SY symmetry direction. This suggests the sample is a \emph{p}-type semiconductor and the CBM lies above the Fermi level thus inaccessible in ARPES measurement, which is consistent with the previous work \cite{arpes2}. In addition, the VBM disperses isotropically, mainly along the $\Gamma$X and $\Gamma$Y symmetry directions, with almost similar in-plane hole effective masses indicating that the in-plane anisotropy in the electrical transport reported by Pi \textit{et al.}, \cite{arpes5} has other extrinsic origin and anisotropic defects distribution \cite{arpes6} might be an important factor to take into account.

\begin{center}
\bfseries{B. Field-effect transistor characteristics}\\
\end{center}

\begin{figure}[ht]
\centerline{\includegraphics[scale=0.85, clip]{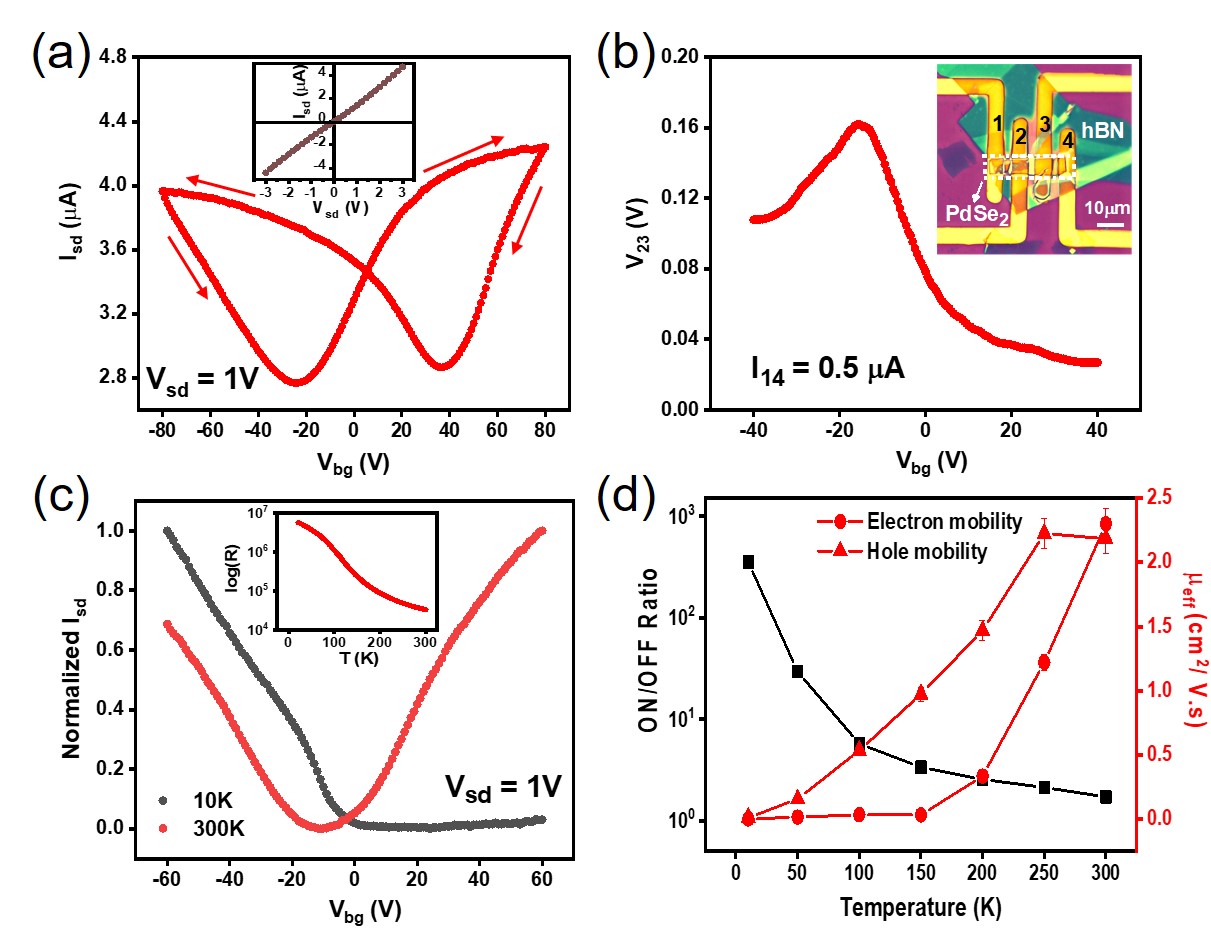}}
\caption{(a) Transfer curve of a typical bulk two-terminal PdSe$_2$ FET illustrating ambipolar transport. The arrow direction represents a clockwise hysteresis in the $I_{sd}$ $vs.$ $V_{bg}$ curve. Inset shows $I_{sd}$ $vs.$ $V_{sd}$ plot at $V_{bg}$=0 V. (b) Transfer curve of a four-terminal current-biased PdSe$_2$ FET with electron mobility 21 cm$^2$/V.s and hole mobility 3.4 cm$^2$/V.s. Inset shows an optical micrograph of a four-probe PdSe$_2$ device embedded with a few-layer hBN flake. (c) Temperature dependent transfer characteristics showing a transition from ambipolar to \emph{p}-dominated transport. The red line indicates the 300 K curve and the black one represents the 10 K curve. Inset shows the resistance $vs.$ temperature plot in logarithmic scale. (d) ON/OFF ratio (black line) of the two-probe PdSe$_2$ transistor and the corresponding field-effect mobilities (red lines) for both electrons and holes versus temperature (Error bar represents 5$\%$ standard deviation). Data points in circles and triangles in the mobility curve represent the electrons and holes, respectively.    
\label{FET-pristine}}
\end{figure}

To investigate the electronic transport properties of pristine and doped samples, PdSe$_2$ FET devices were fabricated in both two-probe and four-probe configurations using optical lithography followed by Cr-Au evaporation. The variation of the source-drain current ($I_{sd}$) as a function of back-gate voltage ($V_{bg}$) with 1V bias ($V_{sd}$) in a two-terminal device with PdSe$_2$ flake, shown in Figure \ref{FET-pristine}a, exhibits ambipolar behavior with a slight asymmetry between electrons and holes. The prominence in the \emph{n}-type characteristics is due to the Se surface vacancies created naturally during the exfoliation from single crystals which causes the excess of electrons in the system. The clockwise hysteresis in $I_{sd}$ $vs.$ $V_{bg}$ curve implies the defect states encountered in the PdSe$_2$ device are static defects \cite{hysteresis}. Inset of Figure \ref{FET-pristine}a shows the linear variation of drain current (I$_{sd}$) with source-drain bias voltage (V$_{sd}$) at $V_{bg}$=0 V indicating that a good ohmic type of contact is formed at the metal junction contrary to the conventional TMDCs (\textit{e.g.} MoS$_2$), thereby reduces the probability of Fermi level pinning effect at the metal-semiconductor junction \cite{pinning1,pinning2}. For the four-probe measurement, the gate-dependence of the voltage between two middle probes with 0.5 $\mu$A current bias is shown in Figure \ref{FET-pristine}b. Inset of Figure \ref{FET-pristine}b illustrates an optical microscopic image of a four-terminal hBN covered bulk PdSe$_2$ device. Although 90\% of the fabricated devices which are exposed in ambient condition show stable ambipolar transfer characteristics with time (more than six months with 1-2\% variation in mobility observed in 16 devices), the PdSe$_2$ device (shown in the inset) was protected by a few-layer (thickness $\sim$20 nm) hexagonal boron nitride (hBN) to resist any surface contamination, if at all there is any (hBN was used to cover the device from the top of the contacts to avoid the presence of hBN tunnel barrier which can be a major issue for hBN layer under the contacts).  A standard schematic of a four-terminal hBN-PdSe$_2$ device in back-gated configuration is depicted in Figure S2a. The two-probe field-effect mobility ($\mu_{eff}$) was calculated from the linear region of the transfer characteristics using the formula,
\begin{equation}
\mu_{eff,2p}=\frac{L}{W C_{ox}} \frac{dI_{sd}}{dV_{bg}} \frac{1}{V_{sd}}
\end{equation}
where \textit{L} and \textit{W} denote the channel length and channel width of the FET device, $C_{ox}$ is the gate capacitance of the SiO$_2$ layer which can be estimated as,
$C_{ox}=\frac {\epsilon_0 \epsilon_r}{d}$, 
\textit{d} being the thickness of the SiO$_2$ layer and the corresponding dielectric constant $\epsilon_r$ = 3.9.
At room temperature the estimated two-probe field-effect mobility ($\mu_{eff}$) for electrons is slightly greater than that of holes (2.3 cm$^2$/V.s for electrons and 1.6 cm$^2$/V.s for holes) and the transistor ON/OFF ratio is $\sim$ 2. Such a small ON/OFF ratio indicates that bulk PdSe$_2$ exhibits low band gap semiconducting behavior. Room temperature four-probe mobility was estimated from the formula,
\begin{equation}
\mu_{eff,4p}=\frac{D}{W C_{ox}} \frac{d(I_{14}/V_{23})}{dV_{bg}}
\end{equation}
where $V_{23}$ is the voltage between two voltage probes separated by distance D and $I_{14}$ is the current flowing from terminal-1 to 4. In four-probe configuration the electron mobility was estimated as 21 cm$^2$/V.s and the corresponding hole mobility was 3.4 cm$^2$/V.s.
 
When the temperature is decreased towards 10 K, the prominence in the \emph{n}-type behavior is suppressed and the FET characteristics becomes more of \emph{p}-type with an ON/OFF ratio $>$ 100 (Figure S2b). This is expected because at low temperature the thermal energy ($k_BT$) is not sufficient to overcome the energy barrier of the Se defects. So the defect states can not take part in conduction. A comparative study of the temperature-dependent transfer characteristics is depicted in Figure \ref{FET-pristine}c. Inset shows the resistance (R) $vs.$ temperature (T) curve in log scale which exhibits a semiconducting nature with a resistance of $\sim$30 $k\Omega$ at room temperature and $\sim$5 M$\Omega$ at 10 K. Temperature dependence of the two-probe field-effect mobility as well as transistor ON/OFF ratio are shown in Figure \ref{FET-pristine}d which clearly justifies that at room temperature electron mobility is higher than hole mobility and as the temperature decreases hole mobility dominates. The transport band-gap ($E_g$) was extracted from an Arrhenius plot of the minimum source-drain current ($I_{sd,min}$) $vs.$ temperature (Figure S2c) by using the formula \cite{transport gap},
\begin{equation}
I_{sd,min} \propto exp(-E_g/2k_BT)
\end{equation}
where $k_B$ is the Boltzmann constant. The extracted value of the band-gap is 0.09 eV. The existence of trap-states usually underestimates the band-gap calculated from this method.
  
\begin{figure}[ht]
\centerline{\includegraphics[scale=0.85, clip]{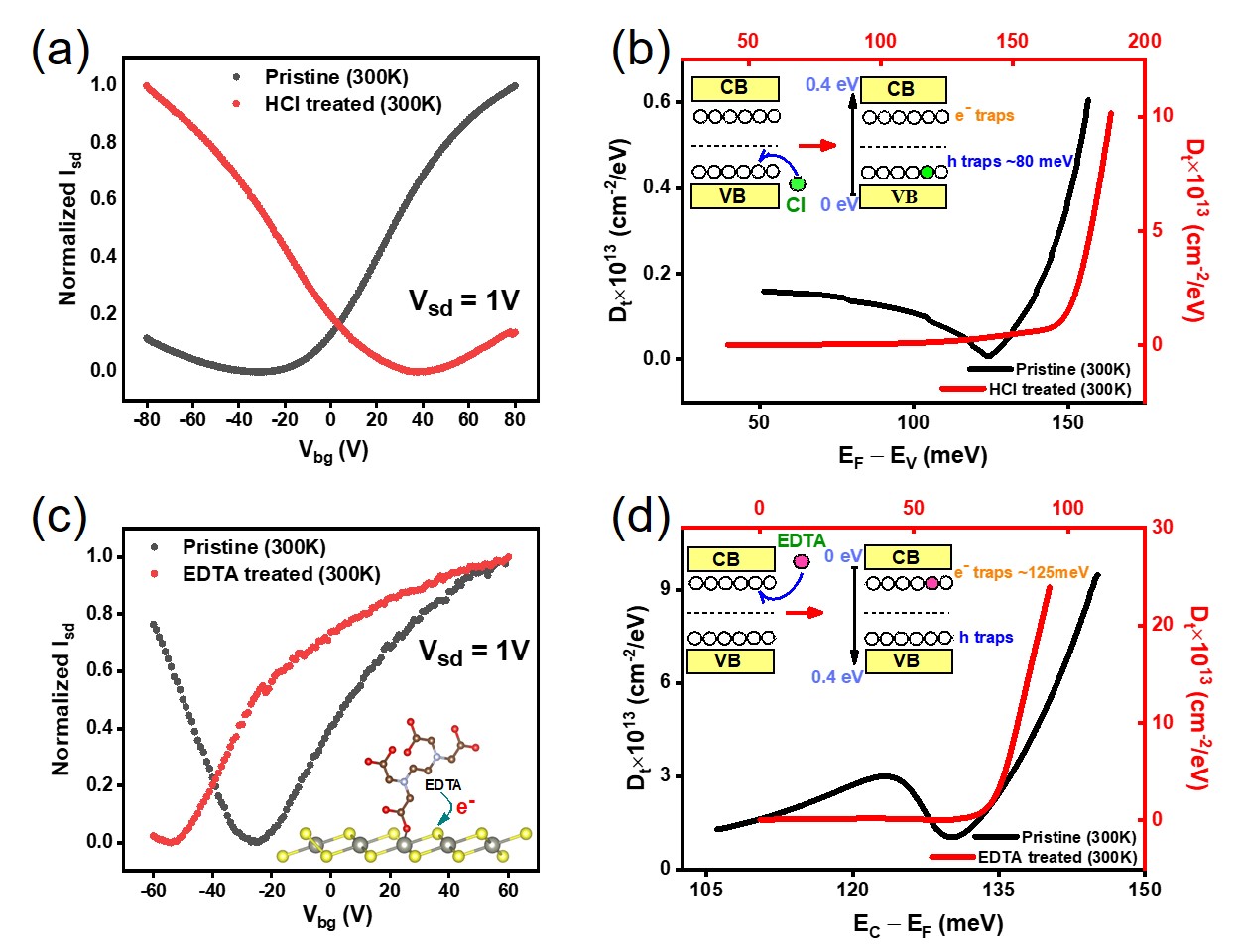}}
\caption{(a) Transfer characteristics corresponding to pristine (black line) and HCl treated (red line) two-terminal PdSe$_2$ device showing a hole-dominated transport after HCl treatment. (b) Density of defect states for pristine (black) as well as HCl treated (red) device illustrating a decrease in the defect states at the valence band side when Cl atoms are attached with the vacancy sites. Inset shows a schematic of repairing the empty hole-traps by Cl treatment. (c) Transfer characteristics corresponding to pristine (black line) and EDTA treated (red line) two-terminal PdSe$_2$ device showing a more electron-dominated transport after EDTA treatment. Inset shows a schematic of EDTA adsorption on the surface of PdSe$_2$. (d) Density of defect states for pristine (black) as well as EDTA treated (red) device illustrating a decrease in the defect states at the conduction band side when EDTAs are attached with the vacancy sites. Inset shows a schematic representing the EDTA treatment of the empty electron-traps. The dotted line represents the intrinsic Fermi level.
\label{doping}}
\end{figure}

After dipping in the HCl solution, the threshold voltage shifts from -20V to 40V towards positive gate bias and the behavior of the PdSe$_2$ transistor turns out to be more of \emph{p}-type (Figure \ref{doping}a) with the hole mobility increased to 5 cm$^2$/V.s. It is predicted that the un-ionized Cl when chemically adsorbed on surface vacancy sites, draws the electrons from the system because of its high electron affinity and thereby, increases the hole carrier injection. At low temperature, the electron transport is almost negligible and the threshold point gets a huge positive shift as shown in Figure S3a. ON/OFF ratio and field-effect hole mobility as a function of temperature for an HCl treated device are depicted in Figure S3b. 

On the other hand, EDTA$^{4-}$ ion acts as a strong electron donor and usually forms metal complexes via four carboxylate and two amine groups in aqueous solution. EDTA is attached by the COO- bond with the Pd atoms around the Se vacancy sites. As an electron donor EDTA repairs the Se vacancies and corresponds to a much prominent \emph{n}-type character with a more negative shift of threshold point (from -25V to -54V) as shown in Figure \ref{doping}c. At low temperature contribution from the defect states is negligible, thereby hole transport gets enhanced (Figure S3c). The ON/OFF ratio and field-effect mobilities as a function of temperature for an EDTA treated device are depicted in Figure S3d. 

EDTA and HCl treated devices were observed with time to check their stability. Notably, the behavior of EDTA doped PdSe$_2$ FET has no such significant change with time as shown in Figure S4a. In case of HCl treated device, the threshold gate voltage suffers a backward shift (from $\sim$40V to $\sim$20V) with time as shown in Figure S4b and the change in hole mobility is around 5-8\%. This nominal change may be a consequence of the air-exposure effect.

\begin{center}
\bfseries{C. Density of defect states}\\
\end{center}

The hysteresis observed in the gate-dependent transfer characteristics as shown in Figure \ref{FET-pristine}a corresponds to the trapped charges encountered by the defect states in the device \cite{trap-1,trap-2,trap-3}. A simple correlation between the width of the hysteresis and the density of the defect states can be illustrated by differentiating the trapped charge concentration ($\Delta N_t$) with respect to the corresponding Fermi level \cite{density of defect states-MoS2}. $\Delta N_t$ can be calculated from the formula : $\Delta N_t$ = $C_{ox} \Delta V_{bg}/q$, where $\Delta V_{bg}$ is basically the width of the hysteresis. Density of defect states ($D_{t}$) can now be estimated from $|\frac{d\Delta N_t}{dE_F}|$ (see supplementary for detail calculation). Figure \ref{doping}b represents the density of defect states at 300 K of the pristine as well as HCl doped device towards the valence band side. Likewise, Figure \ref{doping}d represents the same for pristine as well as EDTA doped device towards the conduction band side. Chlorine draws the electrons from the system and injects holes, thus reduces the defect-induced electron conduction. The injected hole carriers fill the empty shallow traps near the valence band side ($\sim$ 50-100 meV from valence band edge) and acts as an acceptor level for \emph{p}-type conduction resulting in a decrease in the density of defect states in the vicinity of valence band edge (Inset shows the schematic of repairing the defect states by Cl treatment taking 80 meV as a reference energy point of trap states). In a similar manner, when the device is processed through EDTA treatment, EDTA donates excess electrons in the channel and the electron traps within the band gap ($\sim$ 125 meV from conduction band edge) get filled resulting in the enhancement of electron transport (Inset shows the schematic of repairing the defect states by EDTA treatment taking 125 meV as a reference energy point of trap states). This justifies that Cl or EDTA repairs the Se vacancies and thereby causes a decrease in the number of defect states (here the band gap of pristine device is considered to be 0.4 eV as estimated from DFT calculation).

\begin{center}
\bfseries{D. Density functional theory}\\
\end{center}

\begin{figure}[ht]
\centerline{\includegraphics[scale=0.7, clip]{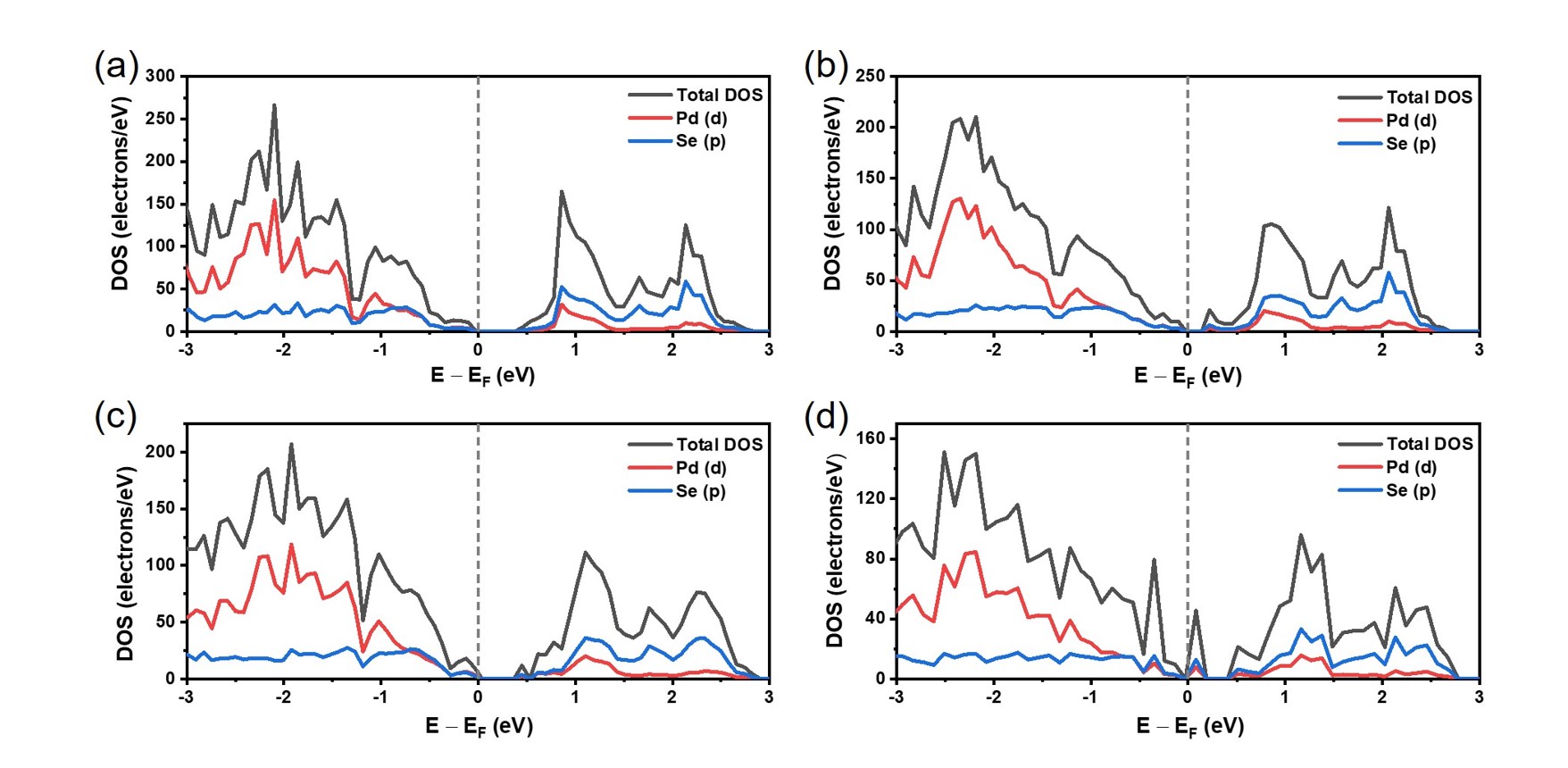}}
\caption{First principle calculations: Density of states of (a) pristine bulk PdSe$_2$, (b) PdSe$_2$ with a single Se vacancy defect showing the defect states arise at the conduction band side, (c) Cl-doped PdSe$_2$ and (d) EDTA-treated PdSe$_2$.   
\label{DFT}}
\end{figure}

The simulated structure parameters of the fully relaxed geometry obtained from Density functional theory calculation are $a=5.77$ {\AA}, $b=5.93$ {\AA}, $c=8.42$ {\AA}, which shows a good agreement with the experimental parameters obtained from single crystal XRD \cite{structure-PdSe2}. The electronic band structure shown in Figure \ref{arpes}b of pristine bulk PdSe$_2$ exhibits a 0.4 eV indirect band gap. According to the calculated density of states (Figure \ref{DFT}a), there is a transition from Pd $4d$ states (valence band side) to Se $4p$ states (conduction band side) which indicates that charge transfer occurs from Pd to Se. This was also verified by bader charge analysis using VASP which shows 0.03e amount of charge is transferred from Pd to each Se. Phonon modes are simulated for pristine sample depicted elaborately in supplementary. Figure \ref{DFT}b displays the in-gap states arise closed to the conduction band when Se surface vacancies are imported and thereby causes a certain decrease in the band gap (0.16 eV). The defect state thus generated corroborates with the slight dominance in electron conduction at room temperature. Cl-doping at the Se vacancy sites repairs the gap-states and modifies the density of states as shown in Figure \ref{DFT}c which implies a shifting of Fermi level towards valence band and thereby amplifies the \emph{p}-type character as obtained from FET measurement. On the other hand, after the adsorption of EDTA molecule some impurity states arise between the Fermi level and conduction band edge (Figure \ref{DFT}d) which well agrees with the electron-enhanced transport in EDTA-treated PdSe$_2$ channel.

\section{Conclusions}
\label{sec:conclusion} 

In summary, PdSe$_2$ single crystals are synthesized successfully with desirable stoichiometry. The ambipolar behaviour of PdSe$_2$ microelectronic devices with low ON/OFF ratio offers moderate electron mobility that may be enhanced in its monolayer analogue. Also, we demonstrate the tunability in the electronic transport properties of bulk PdSe$_2$ layer as a result of chemical doping. The dominance of the electron-transport in the pristine device is arrested when the device is processed by HCl. On the other hand, EDTA/Na$_2$ salt solution acts as an \emph{n}-type dopant and almost suppresses the hole-transport. The phonon modes for pristine, as well as doped samples, are identified using Raman spectroscopy experiment. The effect of vacancy defects as well as chemical doping, as observed in the transfer charateristics, are also supported by the first principle DFT calculations. A simplified analytic approach is considered to calculate the density of defect states which also clarifies that Cl and EDTA occupy the hole traps and electron traps, respectively, and resulting in a decrease of the number of defect states. In addition, temperature dependent electronic properties of PdSe$_2$ depicts the change of the nearly symmetrical ambipolar characteristics at room temperature to almost \emph{p}-type behavior at low temperature which is also confirmed by the low temperature $\mu$-ARPES spectra. Our results illustrate that PdSe$_2$ can be a promising candidate in fabricating single transistor logic gates in the form of dual-gated complementary metal-oxide-semiconductor (CMOS). Also, the tunable ambipolarity would be beneficial in the field of hyperscaling and multivalued logic circuitry.


\section{Acknowledgments}
The financial supports from IACS, DST-INSPIRE and CSIR-UGC are greatly acknowledged. TK would like to thank Prof. Debashree Ghosh along with her lab members, Mr. Sudipta Biswas and Mr. Soumya Mondal for computational help. TK appreciates Mr. Satyabrata Bera, Mr. Soumik Das, Mr. Sourav Mondal, Mr. Sanjib Naskar and Ms. Anwesha Datta for the experimental support. She also acknowledges the furnace facility of Dr. Mintu Mondal, procured under DST-SERB grant No. SRG/2019/000674. TK is also thankful to IACS cluster facility for the computational study reported in this paper. SD acknowledges the financial support from DST-SERB grant No. CRG/2021/004334. SD also acknowledges support (lithography facility) from the Technical Research Centre (TRC), IACS, Kolkata. The research leading to micro-ARPES results in Elettra under Proposal No. 20220474 has been supported by a grant from the Italian Ministry of Foreign Affairs and International Cooperation and the Indian Department of Science \& Technology.
\newline
S.D., and T.K. conceived the project and designed the experiments. T.K., B.D., and R.P. prepared the samples and performed their initial characterization together with S.M.  M.P., S.M., and T.K. carried out the device fabrication and modification. T.K., and B.D. performed the electrical measurements and analyzed the data. S.D., and B.P. developed the analytical model, with contributions from B.D. S.K.M., B.D. and S.D. conceived the ARPES measurements. M.K., and A.B. provided the technical help during the beamtime. All DFT calculations were performed by T.K. with the close discussions with S.D. All authors discussed the results and actively commented on the manuscript written by T. K. and S.D. with the the input from S.K.M. 

\clearpage

\clearpage

\end {document}


\maketitle
\clearpage
\section{Density of defect states}
The gate transfer characteristics of a PdSe$_2$ field-effect transistor shows an ambipolar behaviour with a prominent clockwise hysteresis (Figure 3a). When the transistor slowly switches from zero to positive gate voltage, the Fermi level ($E_F$) starts shifting towards the conduction band edge and therefore, the defect states present in between will be filled with electrons. After that, when it comes back to zero from the highest gate voltage given, all the trapped electrons will be de-trapped. The trapping and de-trapping of charge carriers by the defect states yield the hysteresis observed in the device \cite{density of defect states-MoS2}. Number of trapped charges can be calculated from, 
\begin{equation}
\Delta N_t = C_{ox} \Delta V_{bg} /q, 
\end{equation}
where $C_{ox}$ is the gate-dielectric capacitance, $\Delta V_{bg}$ represents the width of the hysteresis and q is the unit charge. The Fermi energy level, 
\begin{equation}
E_F = E_C + k_B T ln(n/N_c), 
\end{equation}
where, $E_C$ denotes the conduction band minimum, n is the electron concentration found from the measured drain current, $N_C = \frac{g_c m^* k_B T}{\pi \hbar^2}$, $g_c$ being the valley degeneracy and $m^*$ the effective mass of electron. 
If we assume the Fermi level shifts from $E_F$ to $E_F'$ towards the positive gate bias, then the number of trapped electrons in this regime in terms of the density of defect states ($D_{t}$) is given by,
\begin{equation}
\Delta N_t = \int_{E_V}^{E_C} D_{t}(E)f(E-E_F')\,dE\ - \int_{E_V}^{E_C} D_{t}(E)f(E-E_F)\,dE\ 
\end{equation}
Considering the probability of filled states below the Fermi level is 1 and that above the Fermi level is 0, eq.(3) can be reconstructed as,
\begin{equation}
\Delta N_t = \int_{E_V}^{E_F'} D_{t}(E)\,dE\ - \int_{E_V}^{E_F} D_{t}(E)\,dE\ 
\end{equation}   
Now $D_{t}$ in between Fermi level and conduction band edge can be found simply by differentiating the trapped electron concentration with respect to the Fermi level ($|d\Delta N_t/dE_F|$). In a similar manner, the density of defect states in between valence band edge and Fermi level can be estimated by differentiating the trapped hole concentration with respect to the Fermi level.

\section{Phonon modes}
\renewcommand{\thesection}{S\arabic{section}}
Raman spectroscopy was carried out using a LASER excitation of wavelength 532 nm. The resulting unpolarized spectrum (Figure \ref{raman}a) of pristine bulk PdSe$_2$ exhibits six vibrational modes (3$A_g$ and 3$B_{1g}$) appeared around 144 $cm^{-1}$, 206 $cm^{-1}$, 222 $cm^{-1}$, 257 $cm^{-1}$ and 267 $cm^{-1}$ where the first one is a mixed mode of $A_g^1$ and $B_{1g}^1$. These raman modes predominantly involve the vibration of Se atoms \cite{PdSe2} (Figure \ref{raman}b). An exfoliated bulk pristine PdSe$_2$ flake was assigned for Raman spectroscopy with time to check the air-exposure effect which is shown in Figure \ref{raman}a. Interestingly, the relative intensity of the $A_g$ modes decreases and that of $B_{1g}$ modes increases with time and all the peaks slightly shift towards the higher frequency. Also the weak $B_{1g}^3$ mode that was almost suppressed by the nearby strong $A_g^3$ peak in pristine sample, becomes prominent when PdSe$_2$ flakes were exposed in air. The peak intensity variation and shifting of the Raman modes in oxygen-exposure are basically due to the chemisorption of oxygen molecule at Se surface vacancy sites which causes the formation of PdO$_2$ \cite{npj-defect}.

Raman spectrum of pristine bulk PdSe$_2$ was simulated as shown in Figure \ref{raman}c. The theoretical spectrum exhibits six Raman modes with the intensity peaks appeared around 139 $cm^{-1}$, 140 $cm^{-1}$, 198 $cm^{-1}$, 217 $cm^{-1}$, 243 $cm^{-1}$ and 253 $cm^{-1}$. The peak positions differ from the experimental ones by approximately 5-15 $cm^{-1}$. Phonon band structure and density of states shown in Figure \ref{raman}d were also calculated using density functional perturbation theory. The frequencies of the $\Gamma$ point phonon modes have a good match with that of the experimental Raman modes.

After dipping in 11M HCl solution, the Raman modes are slightly shifted towards higher frequency (Figure \ref{raman-doping}a) which is originated from the decrease in bond length between Pd and Cl around the Se vacancies. On the other hand, Raman spectrum of EDTA treated PdSe$_2$ shown in Figure \ref{raman-doping}b depicts that after EDTA treatment the relative intensity of the $A_g$ modes decreases, the weak $B_{1g}^3$ mode becomes prominent as well as the mixed mode of $A_g^1$ and $B_{1g}^1$ can be separately identified. This suggests that \textit{n}-doping corresponds to strong electron-phonon interaction especially for A$_g$ modes \cite{raman-prb} leading to a decrease in the peak intensity. 

\clearpage

\begin{figure}[ht]
\centerline{\includegraphics[scale=0.9, clip]{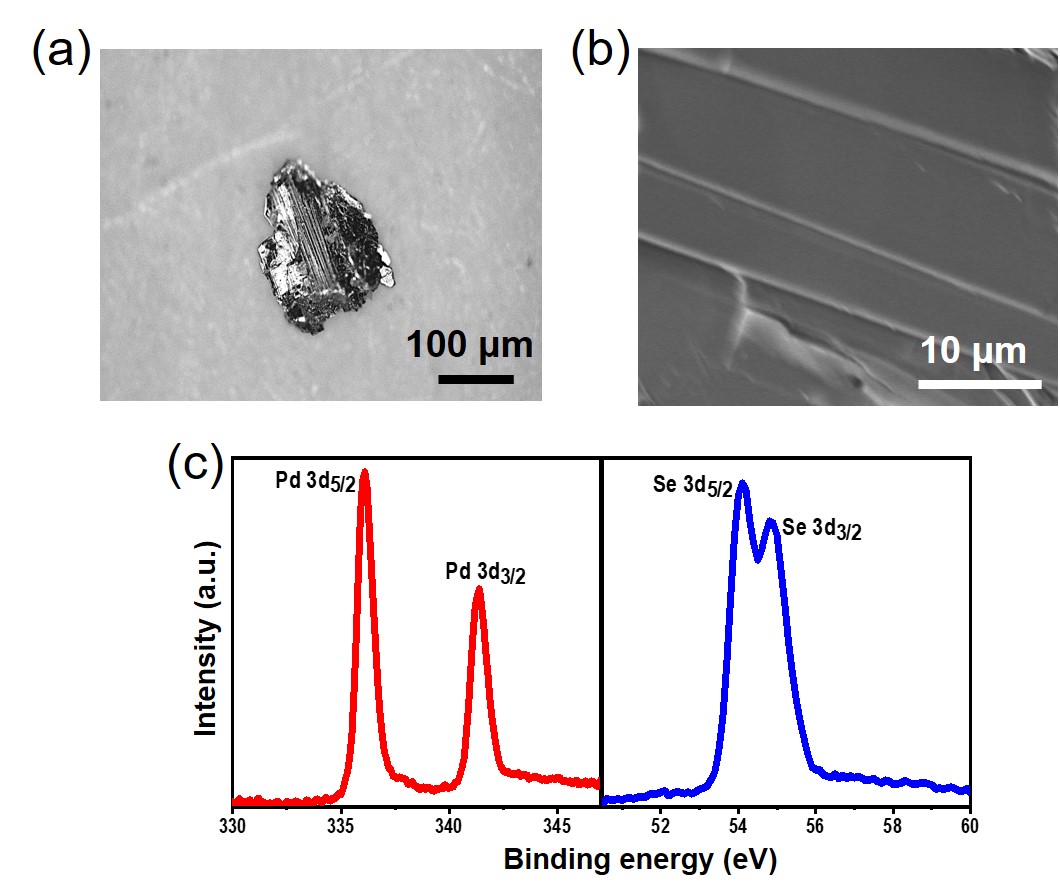}}
\renewcommand{\thefigure}{S\arabic{figure}}
\caption{(a) Optical micrograph of a PdSe$_2$ single crystal. (b) FESEM image of a PdSe$_2$ crystal. (c) Core level X-ray photoelectron spectra of Pd $3d$ and Se $3d$ orbitals.
\label{char}}
\end{figure}

\begin{figure}[ht]
\centerline{\includegraphics[scale=0.62, clip]{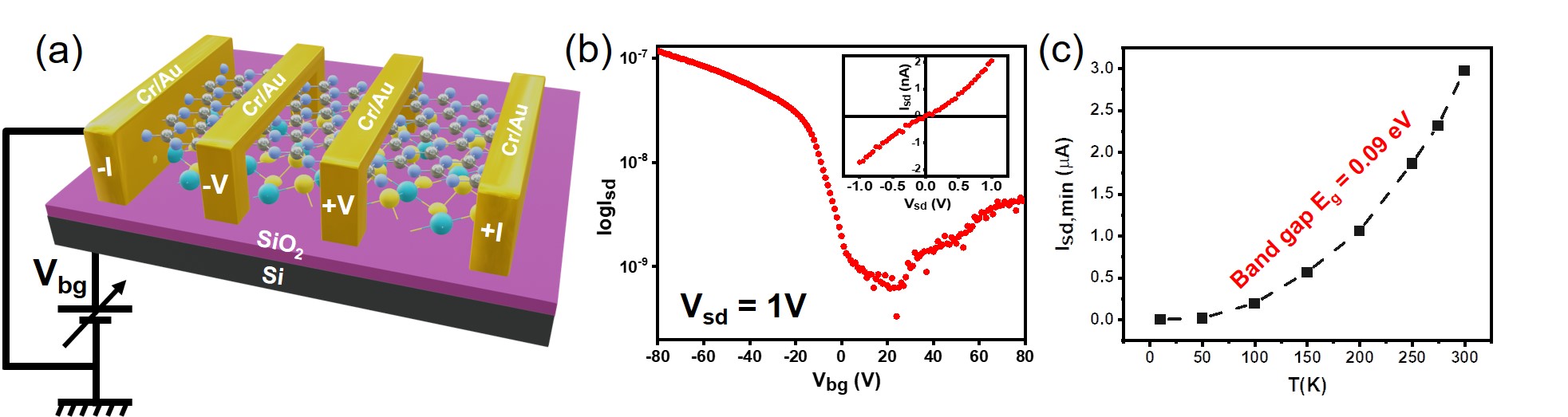}}
\renewcommand{\thefigure}{S\arabic{figure}}
\caption{(a) Schematic representation of a four-terminal back-gated device of PdSe$_2$/hBN heterostructure. (b) Low temperature (10 K) gate-transfer characteristics in log scale of a pristine two-terminal PdSe$_2$ device with 1V bias showing a hole-dominated transport with an ON/OFF ratio $>$ 100. Inset shows the $I_{sd}$ $vs.$ $V_{sd}$ curve at 10 K associated with a zero gate voltage. (c) Minimum drain current plotted as a function of temperature from which extracted value of the transport band-gap is 0.09 eV. 
\label{FET-pristine}}
\end{figure}

\begin{figure}[ht]
\centerline{\includegraphics[scale=0.85, clip]{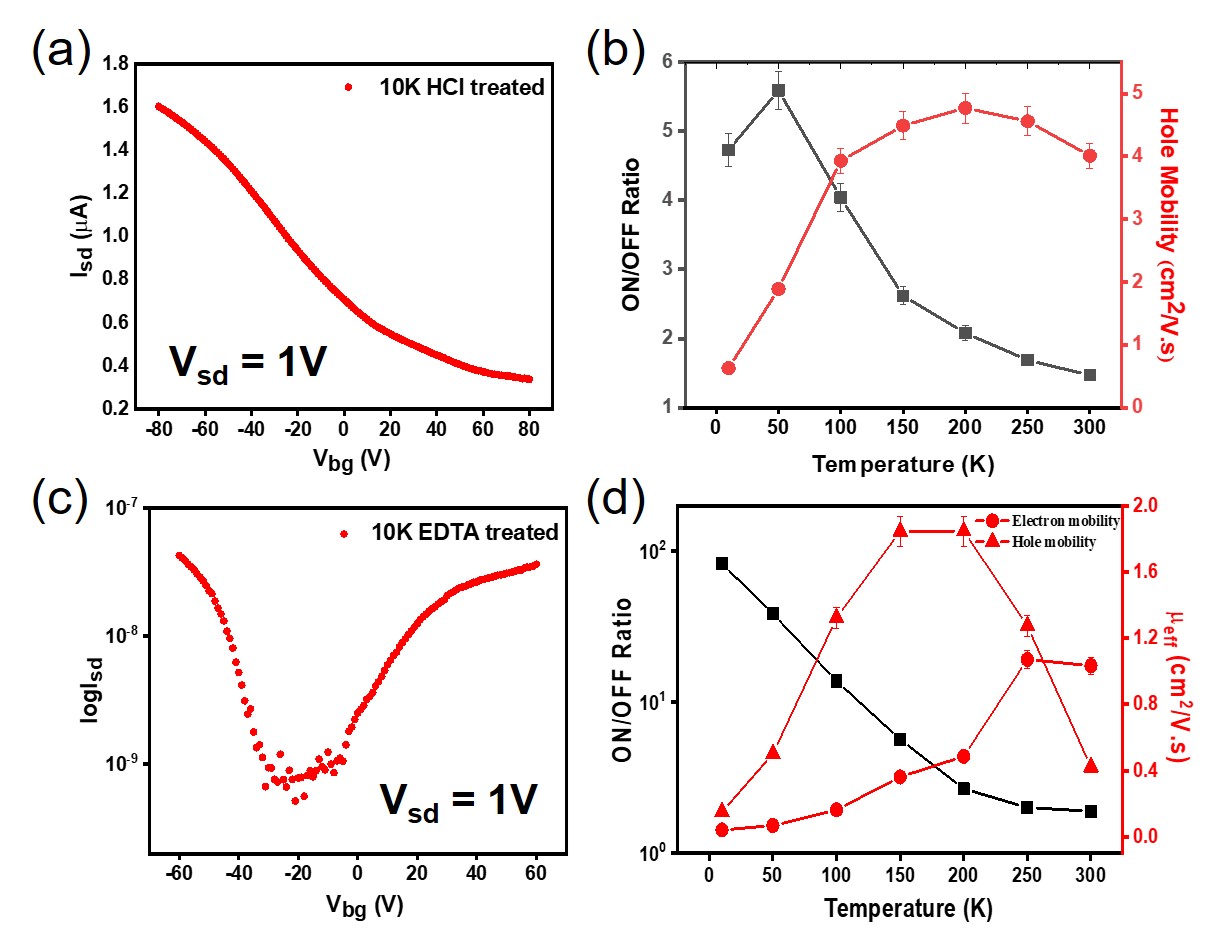}}
\renewcommand{\thefigure}{S\arabic{figure}}
\caption{Low temperature (10 K) gate-transfer characteristics of an (a) HCl, (c) EDTA doped PdSe$_2$ device with 1V bias showing hole-transport enhancement. ON/OFF ratio and mobility as a function of temperature for the (b) HCl and (d) EDTA treated device (Error bar represents 5$\%$ standard deviation).   
\label{doping}}
\end{figure}

\begin{figure}[ht]
\centerline{\includegraphics[scale=0.75, clip]{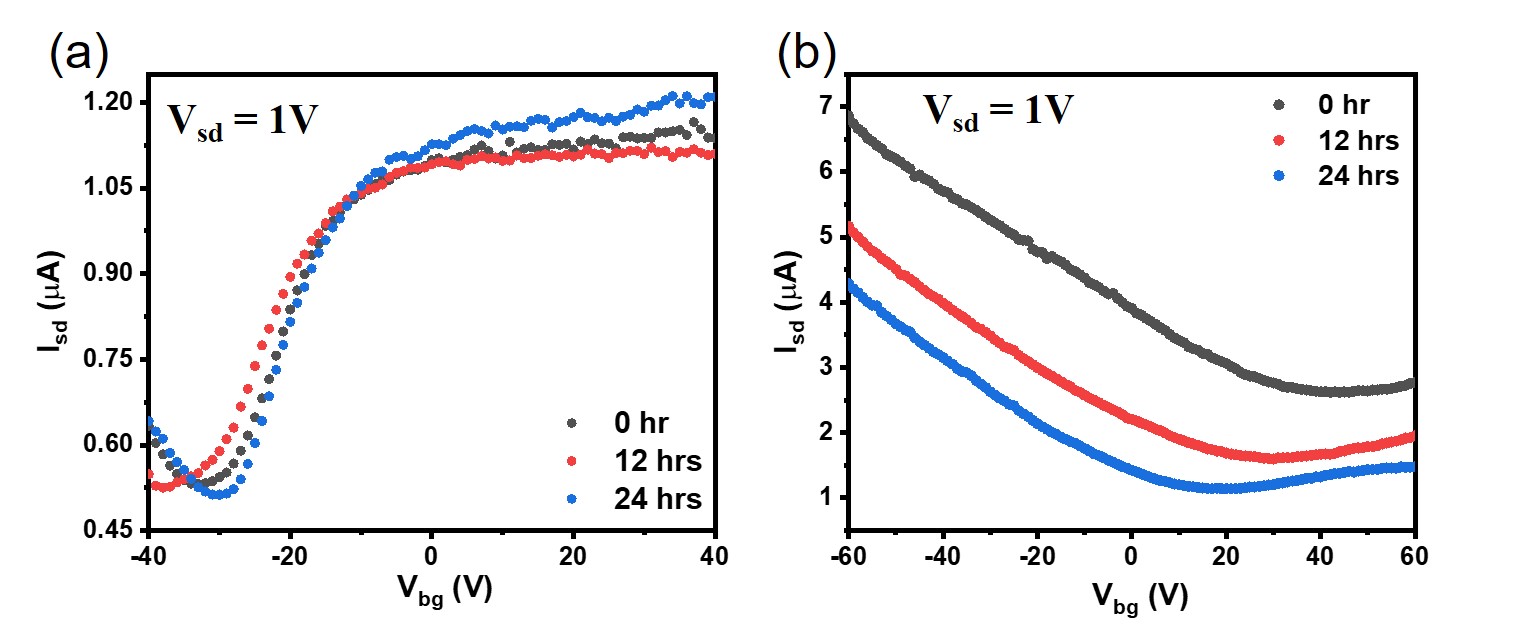}}
\renewcommand{\thefigure}{S\arabic{figure}}
\caption{FET characteristics of air-exposed (a) EDTA treated, (b) HCl treated PdSe$_2$ device with time.
\label{doping stability}}
\end{figure}

\begin{figure}[ht]
\centerline{\includegraphics[scale=0.75, clip]{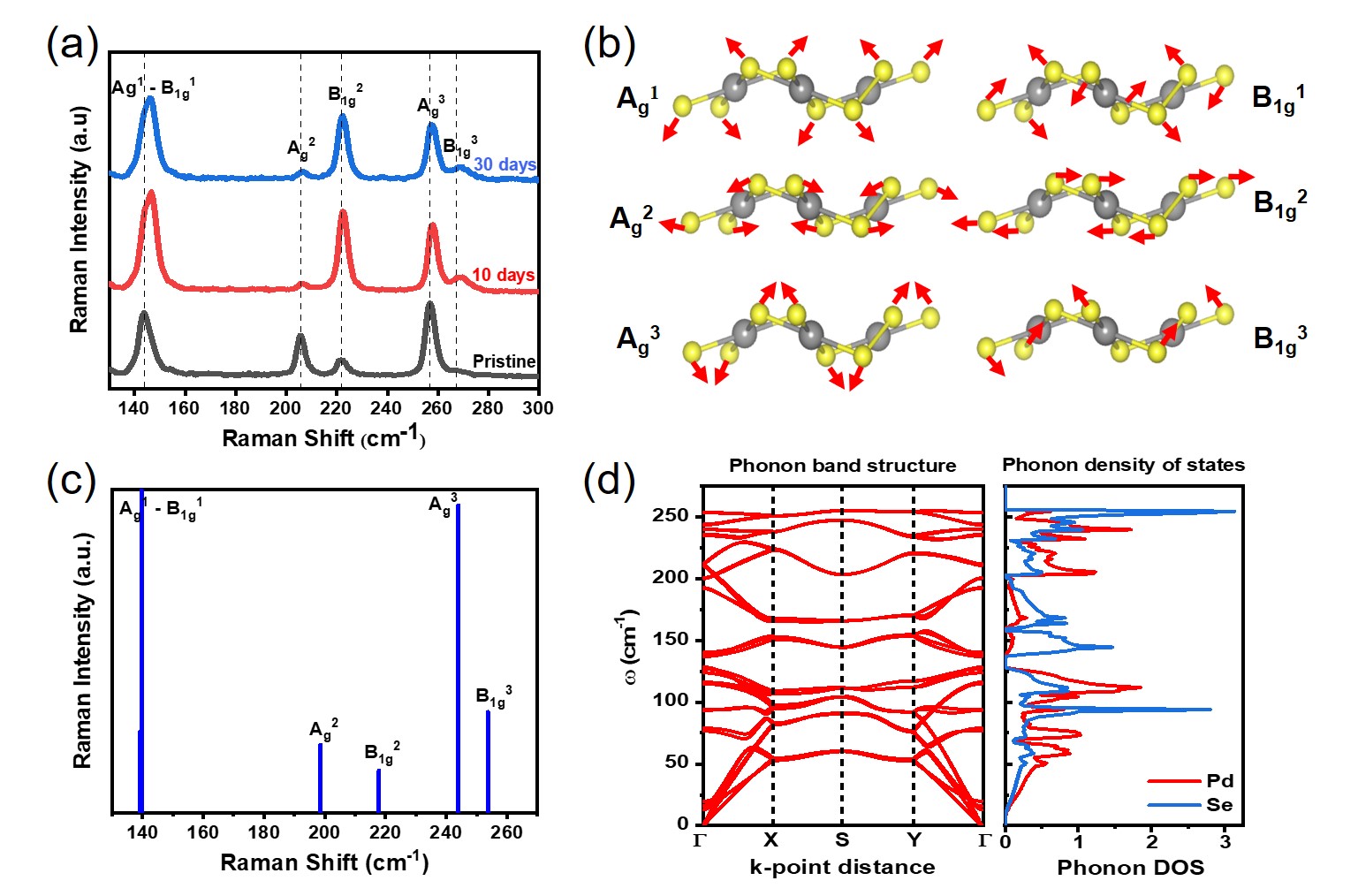}}
\renewcommand{\thefigure}{S\arabic{figure}}
\caption{(a) Unpolarized Raman spectrum of a bulk PdSe$_2$ flake immediately after exfoliation and after 10 days as well as 30 days of exposure in air. (b) Atomic vibrations of the six Raman modes depicted by the red arrows showing that the appeared modes predominantly involve the vibration of Se atoms. (c) Simulated Raman modes, (d) Phonon band structure and density of states for bulk PdSe$_2$ crystal calculated using DFPT approach.
\label{raman}}
\end{figure}

\begin{figure}[ht]
\centerline{\includegraphics[scale=0.85, clip]{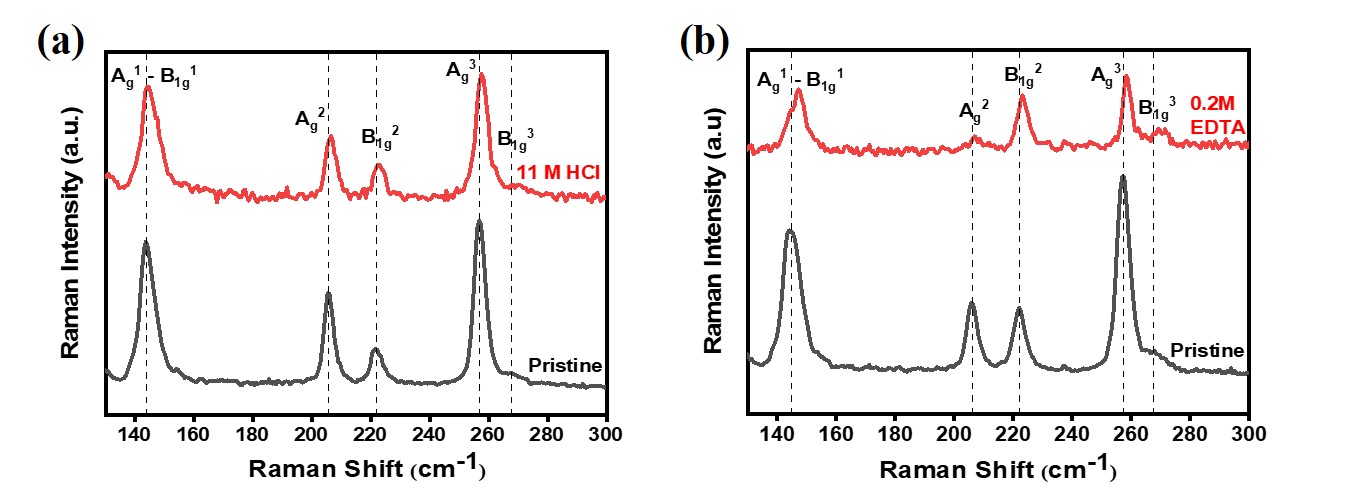}}
\renewcommand{\thefigure}{S\arabic{figure}}
\caption{Raman spectrum of an (a) HCl (b) EDTA doped PdSe$_2$ flake showing a slight blue shift by HCl treatment and a certain decrease in the relative intensity of $A_g$ modes as well as blue shift of the overall spectrum by EDTA treatment.
\label{raman-doping}}
\end{figure}